\documentstyle[prl,aps,epsfig]{revtex}
\begin{document}
\draft
%
%%%%%%%%%%
%
%     Title Page   
%\year 1999 \month  9 \day 14 
%
%%%%%%%%%%
%
\title{ A New Vortex Solution for Two-Component Nonlinear Schr{\"o}dinger 
Equation in Anisotropic Optical Media}
\author{Hiroshi Kuratsuji${}^1$, Shouhei Kakigi${}^1$
 and Hiroyuki Yabu${}^2$}
\address{${}^1$Department of Physics, Ritsumeikan University-BKC, 
Kusatsu City, Shiga 525--8577, Japan\\ 
${}^2$Department of Physics, Tokyo Metropolitan University, 
Hachioji City, Tokyo 192--0397, Japan}
\date{\today}
\maketitle
\begin{abstract} 
A theoretical study is given of a new type of optical vortex in 
nonlinear anisotropic media. This is realized as a special solution 
of the two-component non-linear Schr\"{o}dinger equation. 
The vortex is inherent in the spin texture that is caused by 
an anisotropy of dielectric tensor, for which a role of spin 
is played by the Stokes vector (or pseudo-spin). By using 
the effective Lagrangian for the pseudo-spin field, 
we give an explicit form for the vortex solution for the 
case of two types of optical anisotropy; that is, nonlinear counterpart 
of birefringence giving rise to the Faraday and Cotton-Mouton effects. 
We also examine the evolution equation of the new vortex 
with respect to the propagation direction. 
\end{abstract}

Nonlinear optics has been a major subject in physics for long 
time \cite{Chiao}. The main interest is focussed on particular 
solutions of the basic non-linear equation that governs the light 
field or electromagnetic field. Among others, remarkable is an 
optical vortex, the existence of which has been early suggested in 
\cite{Chiao}. Recently the detailed study has been carried out 
from both of theoretical and experimental point of view 
\cite{Snyder1,Swartz}. More recently experimental verification 
has been also given for the multi-vortices \cite{Rozas}. 
The basic idea of the optical vortex follows an analogy with the 
superfluid vortex that is described by the complex 
order parameter for bose fluid\cite{Swartz}, namely, 
the equation for the light field, which is known as 
the nonlinear Schr\"{o}dinger equation (NLS), is very similar to 
the Pitaevki equation. Thus it is natural to expect the occurrence of 
the optical counterpart of the superfluid vortex. 

The purpose of this letter is to explore a possible new type of 
vortex in nonlinear and anisotropic media that are characterized by 
a variant of birefringence. We call this new type vortex an  
``optical spin vortex''. Our starting 
equation is the two-component NLS. The two-component 
NLS has been recently used for exploring an object of 
dynamic soliton \cite{Snyder2}, which is the 
two modes induced waveguide leading to the soliton polarization 
dynamics. This work shares partly the basic idea with the present 
attempt in the point that two component NLS naturally incorporates the 
polarization state of light. Indeed the concept of polarization 
plays an important role in modern optics, especially 
in crystal optics \cite{Born,Landau}. The quantity describing 
the polarization state is realized by the Stokes parameters, 
which forms a pseudo-spin and 
is geometrically described by a point on the Poincar\'{e} sphere. 
Thus the two component NLS can be written in terms of the field 
of pseudo-spin, which is naturally achieved by introducing 
the effective ``Lagrangian'' of fluid dynamical form. 
The use of the effective Lagrangian gives us a direct access 
to the analysis of the optical spin vortex.
The main thrust is twofold: First we give an explicit form 
for the vortex solution by adopting some specific nonlinear 
birefringence; the concrete form is given by the 
nonlinear counterpart of the one causing 
the Faraday and Cotton-Mouton effects. 
This will be carried out by numerical way. We next examine the evolutional 
behavior of a vortex with respect to the direction of 
the wave propagation. 

\paragraph*{Two component nonlinear Schr\"{o}dinger equation.---}
First we derive the two-component NLS for the light wave traveling 
through anisotropic media. 
The procedure follows the one developed in a recent paper \cite{Kakigi}. 
Suppose that the electromagnetic wave of wave vector $k$ travels 
in the direction of $z$ with the dielectric tensor $\hat \epsilon$. 
The nonlinear nature of media implies that $\hat \epsilon$ has a 
field dependence in nonlinear form, the explicit form of which 
will be given later. Further we assume that $\hat\epsilon$ 
varies slowly compared with the wavenumber $k$. The $z$-axis is chosen 
as a principal axis of the dielectric tensor, namely, 
the axis corresponding to 
one of the eigenvalues of the dielectric tensor. 
In this geometry, $\hat \epsilon$ is taken to be $2 \times 2$ matrix. 
Let us consider the field equation for the displacement field 
${\bf D}$, which is reduced from the Maxwell equation 
\cite{Snyder2,Landau2}: 
\begin{equation}
{\partial^2 {\bf D} \over \partial z^2} + \nabla^2{\bf D} 
+ \left({\omega \over c}\right)^2\hat\epsilon{\bf D} = 0 
\label{one}
\end{equation}
where $\nabla = \left({\partial\over\partial x},\; 
{\partial\over\partial y}\right)$ and $(x,y)$ denotes 
the coordinate in the plane perpendicular to $z$
axis. Now we put
\begin{equation}
{\bf D}(x, y, z) = {\bf f}(x, y, z)\exp[ikn_0z]
\label{two}
\end{equation}
with $k = {\omega\over c}$, and $n_0 (\equiv\sqrt{\epsilon_0})$ 
means the refractive index for the case as if the medium is 
isotropic. The amplitude ${\bf f}(x, y, z)$ is written as 
${\bf f} = {}^t(f_1, f_2) = f_1 {\bf e}_1 + f_2{\bf e}_2$. We 
assume that ${\bf f}$ is slowly varying function of 
$z$ besides $(x, y)$, and ${\bf e}_1$ and ${\bf e}_2$ 
denotes the basis of linear polarization. By substituting 
(\ref{two}) into (\ref{one}) and noting the slowly varying 
nature of ${\bf f}$ i.e., 
$\left\vert {\partial{\bf f} \over \partial z}\right\vert \ll 
k\vert {\bf f} \vert$, we can derive the equation for 
the amplitude ${\bf f}$, namely, we can only retain the first 
derivative ${\partial{\bf f} \over \partial z}$ as well as the 
Laplacian with respect to $(x,y)$ \cite{Akhmanov}, hence 
\begin{equation}
i\lambda{\partial {\bf f} \over \partial z} + 
\left[{\lambda^2 \over n_0^2}\nabla^2 + \left(\hat {\epsilon} - 
n_0^2\right)\right]{\bf f} = 0
\end{equation}
where $\lambda$ is the wavelength divided by $2\pi$. 
This equation is regarded as a two-state Schr\"{o}dinger equation 
where $\lambda$ just corresponds to the Planck constant and $z$ 
plays a role of time variable. The components $(f_1, f_2)$ couple 
each other to give rise to the change of polarization which is just 
the effect of birefringence governed by a $2\times 2$ matrix 
``potential'' $\hat v = \hat\epsilon - n_0^2$. 
$\hat v$ represents a deviation from the isotropic value 
and it becomes hermitian if the non-absorptive medium is concerned. 
 From the hermiticity, the most general form of $\hat v$ is written as 
\begin{equation}
{\hat v} = \left(
 \begin{array}{cc}
 v_0 + \alpha & \beta + i\gamma \\
 \beta -i\gamma & v_0 -\alpha \end{array}
\right).
\end{equation}
For later convenience, we transform the basis to 
the circular basis instead of the linear polarization 
(${\bf e}_1,\, {\bf e}_2)$, that is,
 ${\bf e_{\pm}} = \left(1/{\sqrt 2}\ \right)({\bf e}_1 \pm i{\bf e}_2)$,
which is written as $({\bf e}_{+},\, {\bf e}_{-})
= T({\bf e}_1, {\bf e}_2)$. Here $T$ is given by $2 \times 2$ 
unitary matrix: 
\begin{equation}
 T = {1 \over \sqrt 2} \left(
 \begin{array}{cc}
 1 & i \\
 1 & -i \end{array}
\right).
\end{equation}
By introducing the wave function as 
$\psi = T{\bf f}= {}^t\left(\psi_1^{*},\, \psi_2^{*}\right)$, 
we have the Schr{\"o}dinger equation for $\psi$: 
\begin{equation}
i\lambda{\partial \psi \over \partial z} = \hat H \psi 
\end{equation}
with the transformed ``Hamiltonian'' 
\begin{equation}
\hat H = T h T^{-1} = -{\lambda^2 \over n_0}\nabla^2 + V.
\end{equation} 
The ``field-dependent'' potential $V$ is written in 
terms of the Pauli spin; $V = v_0\times 1 + \sum_{i=1}^3v_i\sigma_i$. 

\paragraph*{Effective Lagrangian for the pseudo-spin field.---} 
We now introduce the ``quantum'' Lagrangian leading to the 
Schr\"{o}dinger type equation, which is given by 
\begin{equation}
 I = \int \psi^{\dagger}\left(i\lambda{\partial \over \partial z} - 
\hat H'\right)\psi d^2x dz.
\end{equation}
Indeed, the Dirac variation equation $\delta I = 0$ recovers 
the Scr\"odinger equation. We write $L_C = \int \psi^{\dagger}
i\lambda{\partial \over \partial z}
\psi d^2x$ and $H = \int \psi^{\dagger}\hat H'\psi d^2x$ 
with ($H' = T + V'$), which are 
called the canonical term and the Hamiltonian term respectively. 
Here we note that $V'(\psi^{\dagger}, \psi)$ 
differs from $V$ in (7) and some relation holds between 
$V$ and $V'$, namely, 
$V = V' + \psi^{\dagger}{\partial V' \over \partial 
\psi^{\dagger}}$ in order to recover the NLS. 
Having defined the Lagrangian 
for the two-component field $\psi$, we rewrite this in terms 
of the Stokes parameters: This is defined as 
$S_i = \psi^{\dagger}\sigma_i\psi, S_0 = \psi^{\dagger}1\psi$ with 
$i = x,y,z$ \cite{Kakigi,Brosseau}. 
We see that the relation $S_0^2 = S_x^2 + S_y^2 + S_z^2$ 
holds, namely, $S_0$ gives the field strength; $S_0 \equiv 
\left\vert{\bf D}\right\vert^2$. Using the spinor representation,
\begin{equation}
\psi_1 =\sqrt{S_0} \cos{\theta \over 2},\;
\psi_2 =\sqrt{S_0} \sin{\theta \over 2}\exp[i\phi],
\end{equation}
we have the polar form for the Stokes vector 
${\bf S} = (S_x,\, S_y,\, S_z) \equiv (S_0\sin\theta\cos\phi,\; 
S_0\sin\theta\sin\phi,\; S_0\cos\theta)$, 
 which forms a pseudo-spin and is pictorially given by 
the point on the Poincar\'{e} sphere. In terms of the angle variables, 
the Lagrangian is written as 
\begin{equation}
L = \int {S_0 \lambda \over 2}\left(1 - \cos\theta\right)
{\partial \phi \over \partial z}d^2x -\left(H_T + \tilde V\right)
\label{three}
\end{equation}
where the potential term $\tilde V$ becomes  
\begin{equation}
\tilde V = \int \left(v_0' + \sum_{i=1}^{3}v_i'S_i\right)d^2x.
\end{equation}
Here $v_0'$ and $v_{i}'$'s are nonlinear functions 
of the field strength $S_0$ as well as the angular functions 
$(\theta,\, \phi)$ and this feature may be required for 
a stability of special solution for the pseudo-spin field. 
The kinetic energy term $H_T$ is given as a sum of three terms; 
$H_T = {\lambda^2 \over n_0 }\int \nabla \psi^{\dagger}\nabla \psi d^2x
\equiv H_1 + \tilde H$ where the first term becomes 
$H_1 = \int {\lambda^2 \over 4S_0 n_0}\left(\nabla S_0\right)^2d^2x$, 
which gives the energy that is needed for space modulation of the 
field strength, and the remaining terms are written as 
\begin{equation}
\tilde H = \int {S_0\lambda^2 \over n_0}\left\{(\nabla\theta)^2 
+ \sin^2{\theta \over 2}(\nabla\phi)^2\right\}d^2x,
\end{equation}
which is separated into two terms; $\tilde H = H_2 + H_3$, 
\begin{eqnarray}
H_2 & = & \int {S_0 \lambda^2 \over 4 n_0}
\{(1-\cos\theta)\nabla\phi\}^2d^2x, \nonumber\\
H_3 & = & \int {S_0 \lambda^2 \over 4n_0}\{(\nabla\theta)^2 
+ \sin^2\theta(\nabla\phi)^2\}d^2x.
\label{pseudo}
\end{eqnarray}
Here if we define the ``velocity field'' 
${\bf v} = (1 -\cos\theta)\nabla\phi$, 
the first term is regarded as fluid kinetic energy 
inherent in spin structure, while the last term represents an 
intrinsic energy for the pseudo-spin which exactly coincides with 
a continuous Heisenberg spin chain \cite{Ono}. 

\paragraph*{Vortex solution and its numerical evaluation.---} 
We are now concerned with getting an explicit form for 
the specific type of solutions, namely, vortex solution for 
the two-component NLS. The solution we want here is a 
``static'' solution, namely, we look for the solution 
that is independent of the variable $z$. 
For this purpose, we consider two types of anisotropy. 

\begin{figure}[htpb]
\begin{minipage}[b]{8.2cm}
\epsfig{file=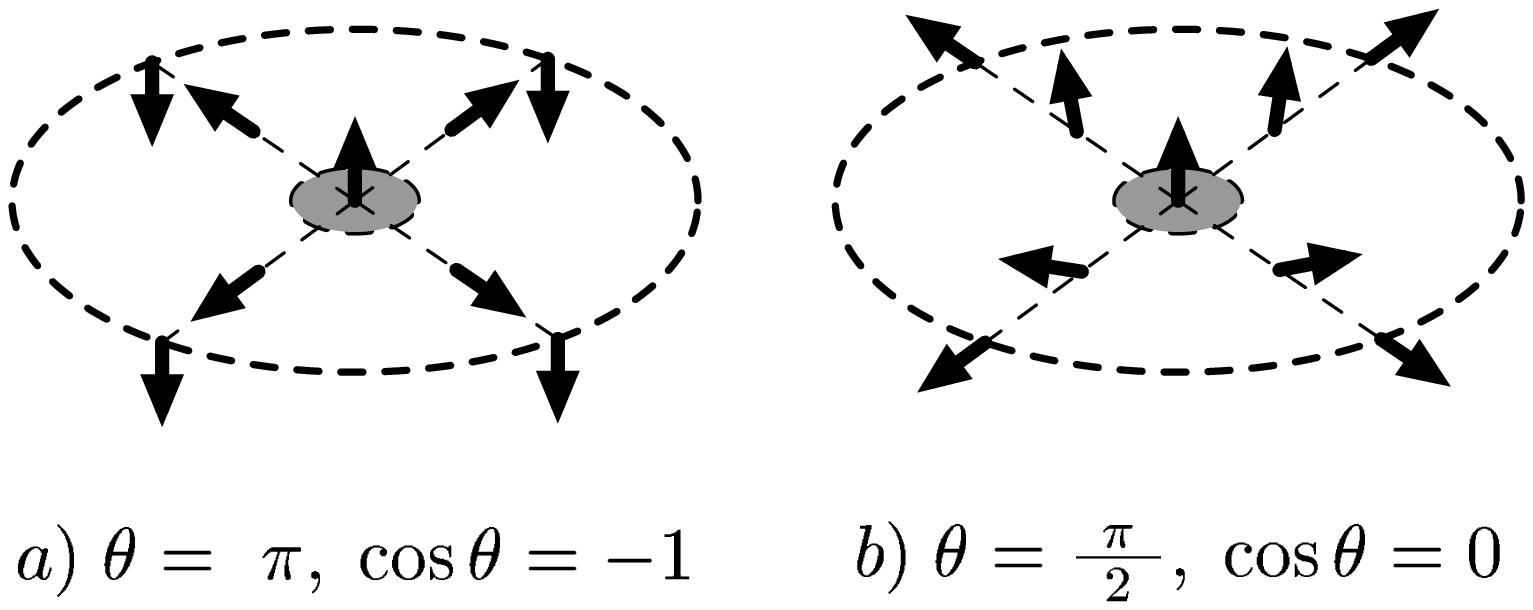,width=8.2cm}
\caption{The profile  of the non-singular vortex; a) The case of 
$\theta(\infty) = \pi$ and b) The case of $\theta(\infty) = {\pi \over 2}$.}
\label{fig1}
\end{minipage}\hfill
\begin{minipage}[b]{8.2cm}
\epsfig{file=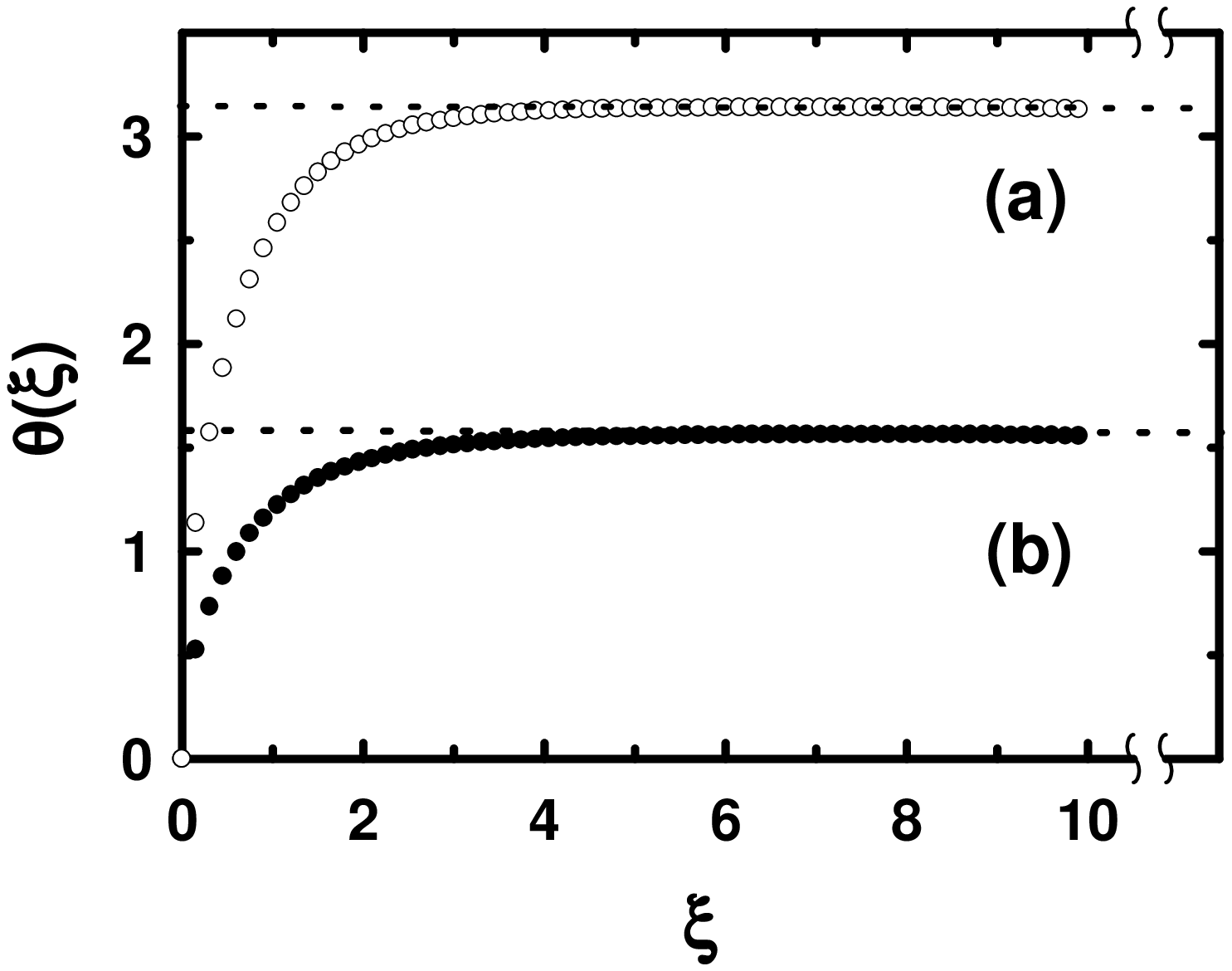,width=8.2cm}
\caption{The profile of the function $\theta(\xi)$ for 
two cases of boundary conditions: (a) $\theta(\infty) = \pi$ 
and (b) $\theta(\infty) = {\pi \over 2}$.}
\label{fig2}
\end{minipage}
\end{figure}

(I) First we adopt the following nonlinear birefringence:
\begin{equation}
 \hat v = \left(
 \begin{array}{cc}
g( \psi_1^{*}\psi_1 - \psi_2^{*}\psi_2) & 0 \\
 0 & -g(\psi_1^{*}\psi_1 - \psi_2^{*}\psi_2) \end{array}
\right) 
\end{equation}
with the positive coupling constant $g$. This may 
be regarded as a nonlinear realization of birefringence that 
causes the Faraday effect. We have the potential 
$\tilde V = \int v'_3S_3d^2x = gS_0^2\int \cos^2\theta d^2x$. 
In what follows, we confine our argument to the case that $S_0$ becomes 
constant. Physically, this corresponds to the constant background field 
with a proper core which is controlled by the profile of the 
angle functions $(\theta,\, \phi)$. 
A static solution for the one vortex is obtained 
by choosing the phase function $\phi = n\tan^{-1}\left({y \over x}\right)$,
with $n = 1,2, \cdots.$ being the winding number, 
together with the profile function $\theta$ that 
is given as a function of the radial variable $r (= \sqrt{
x^2 + y^2})$. Note that such a vortex becomes non-singular, 
namely, the velocity field ${\bf v} = (1 - \cos\theta)$ does not 
bear the singularity due to the 
behavior of $\theta(r)$ near the origin (see below). 
The static Hamiltonian is thus written in terms of 
the field $\theta(r)$:
\begin{equation}
H' = {S_0\lambda^2 \over 4n_0} 
\int\left[\left\{\left({d\theta \over dr}\right)^2 
+ {n^2\over r^2} \sin^2{\theta\over 2}\right\} + g'\cos^2\theta\right]rdr
\end{equation}
where $g'= {4gn_0 S_0 \over \lambda^2}$. 
The profile function $\theta(r)$ may be derived 
from the extremum of $H'$, namely, the Euler-Lagrange equation 
leads to 
\begin{equation}
{d^2 \theta \over d\xi^2}
+ {1\over \xi}{d\theta \over d\xi} - {n^2\over 4\xi^2}\sin\theta 
       + {1\over 2}\sin2\theta = 0
\end{equation}
where we adopt the scaling of the variable: 
$\xi =\sqrt{g'}r$. In order to examine the behavior of $\theta(\xi)$, 
we need a specific boundary condition 
at $\xi =0$ and $\xi= \infty$. We impose 
$\theta(0)= 0$, whereas at $\xi=\infty$, there are 
two options: a) $\theta(\infty)= \pi$ and b) $\theta(\infty) =\pi/2$. 
If introducing 
the vector ${\bf m}(\xi) \equiv {\bf S}/S_0$, 
we have $m_3(0)=1$ for both cases a), b) and we have 
$m_3(\infty)=-1$ for case a) and $m_3(\infty)=0$ for case b). 
This feature indicates that the pseudo-spin field which directs {\it upward} 
(left-handed circular polarization) at the origin changes to 
the state of {\it downward}(right handed polarization)
 or {\it outward}(linear polarization) with departing from the 
origin [see Fig.\ref{fig1}(a) and (b)]. 
We first consider the behavior near the origin $\xi = 0$, 
for which the differential equation behaves like the Bessel equation, 
so we see $\theta(\xi) \simeq J_{n/2}(\xi)$, which satisfies 
$\theta(0) \simeq 0$. 
%
%%%%%%%%%%%%%%%%%%%%%%%%%%%%%%%%%%%%%%%%%%%%%%%%%%%%%%%%%%%%%%%%%%%%%%
%%                                                                  %%
%%   Put Fig. 1 around here                                         %%
%%                                                                  %%
%%%%%%%%%%%%%%%%%%%%%%%%%%%%%%%%%%%%%%%%%%%%%%%%%%%%%%%%%%%%%%%%%%%%%%
%% \begin{figure}                                                   %%
%% \caption{The profile of the non-singular vortex; a) The case of  %%
%% $\theta(\infty) = \pi$ and b) The case of $\theta(\infty) =      %%
%% {\pi \over 2}$.}                                                 %%
%% \label{fig1}                                                     %%
%% \end{figure}                                                     %%
%%%%%%%%%%%%%%%%%%%%%%%%%%%%%%%%%%%%%%%%%%%%%%%%%%%%%%%%%%%%%%%%%%%%%%
We examine the behavior at $\xi = \infty$. This is simply 
performed by checking the stability for two cases mentioned above: 
(a) and (b). Now for the 
case (b), if putting $\theta(\xi) = {\pi \over 2}
+ \alpha$, with $\alpha$ the infinitesimal deviation, 
then we have the linearized equation  $\alpha'' - \alpha \simeq 0$ 
near $\xi = \infty$, 
which results in $\alpha \simeq \exp[-\xi]$. This means that 
the solution with $\theta(\infty) = {\pi \over 2}$ is stable.
On the other hand, for the case (a) we have $\alpha'' 
+ \alpha \simeq 0$, which gives $\alpha \simeq 
\exp[{\pm}i\xi]$ meaning 
the oscillatory behavior. This simply implies that the solution 
with $\theta(\infty) = \pi$ does not converge to the stable 
solution which means that the case (a) is not relevant. 
%
%%%%%%%%%%%%%%%%%%%%%%%%%%%%%%%%%%%%%%%%%%%%%%%%%%%%%%%%%%%%%%%%%%%%%%
%%                                                                  %%
%%   Put Fig. 2 around here                                         %%
%%                                                                  %%
%%%%%%%%%%%%%%%%%%%%%%%%%%%%%%%%%%%%%%%%%%%%%%%%%%%%%%%%%%%%%%%%%%%%%%
%% \begin{figure}                                                   %%
%% \caption{The profile of the function $\theta(\xi)$               %%
%% for two cases of boundary conditions: (a) $\theta(\infty) = \pi$ %%
%% and (b) $\theta(\infty) = {\pi \over 2}$.}                       %%
%% \label{fig2}                                                     %%
%% \end{figure}                                                     %%
%%%%%%%%%%%%%%%%%%%%%%%%%%%%%%%%%%%%%%%%%%%%%%%%%%%%%%%%%%%%%%%%%%%%%%
Keeping mind of the above general feature, we here 
give a numerical solution of $\theta(\xi)$ in Fig.\ref{fig2}(b). 

(II) We next examine the nonlinear birefringence 
that is governed by the off-diagonal $\hat v$ matrix 
such that 
\begin{equation}
 \hat v = \left(
 \begin{array}{cc}
 0 & g_1 \psi_1^{*}\psi_2 \\
 g_1\psi_2^{*}\psi_1 & 0 \end{array}
\right).
\end{equation}
This matrix may be regarded as a nonlinear counterpart 
of the birefringence that causes the so-called 
Cotton-Mouton effect\cite{Landau}. The corresponding potential energy 
becomes $\tilde V = 
\int\sum_{i=1}^2v_i'S_id^2x = S_0^2\int g_1d^2x - S_0^2\int g_1
\cos^2\theta d^2x$, the first of which is constant and 
should be discarded. Thus the resultant equation for 
the profile function $\theta(r)$ leads to
\begin{equation}
{d^2 \theta \over d\xi^2}
+ {1\over \xi}{d\theta \over d\xi} - {n^2\over 4\xi^2}\sin\theta 
       - {1\over 2}\sin2\theta = 0
\end{equation}
where the scaling variable $\xi = \sqrt{g_1'} r$ with $g_1' 
= {4g_1n_0S_0 \over \lambda^2}$. One should note the ``minus sign'' 
in the last term. Due to this, if applying the same 
procedure in the case (I), we see that the behavior near the origin 
is given by $\theta(\xi) \simeq I_{n/2}(\xi)$, the modified Bessel 
function, which satisfies $\theta(0) = 0$. At infinity, 
we get a stable solution for $\theta(\xi)$ such that 
the boundary condition $\theta(\infty) = \pi$ is 
satisfied. This feature is opposite to the previous case, 
namely, the solution satisfying the boundary condition 
$\theta(\infty) = {\pi \over 2}$ oscillates so it should be 
omitted. The numerical result is also given in Fig.\ref{fig2}(a).

In summary, the vortex solutions in these two cases show up 
quite different behaviors each other, 
due to the difference of nonlinear birefringence. From Fig.2 (a) and (b), 
we can estimate the scaled vortex-core size $\xi_c$ 
from $\xi_c \theta(\infty) =S$
where the area $S$ surrounded by the solution curve and the asymptotic 
line $\theta =\theta(\infty) =\pi$ (for case (a)) or $\pi/2$ (for case (b)). 
The $\xi_c$ is just the mean value for the area $S$. 
Using the numerical results, 
we get 
$\xi_c = S/{\pi \over 2} = 0.714$ for (b) and 
$\xi_c = S/\pi = 0.61$ for (a). 
which are consistent with half the scaled coherent (healing) length 
$\xi_{coh}/2 =r_{coh}/\sqrt{g'} =1/\sqrt{2} \sim 0.7$
The real (unscaled) core-size $r_c$ is given by $r_c =\xi_c/\sqrt{g'}$ 
for both (a) and (b). 
If the characteristic wavelength of light is larger than the core size; 
the core should be detectable: its condition is given by 
$\xi_c < \lambda \sqrt{g'}$. 

\paragraph*{Evolution equation for vortex.---} 
Having demonstrated the explicit form for the 
vortex solution, we now consider the evolutional behavior 
for a single vortex with respect to the propagation direction $z$. 
Following the procedure used in the magnetic vortex \cite{Ono}, 
let us introduce the coordinate of the center of vortex, 
${\bf R}(z) = (X(z), Y(z))$, 
by which the vortex solution is parameterized such that 
$\theta({\bf x} - {\bf R}(z))$ and $\phi({\bf x} - {\bf R}(z))$. 
By using this parametrization, the canonical 
term $L_C$, the first term in (\ref{three}), is written as 
\begin{equation}
 L_C = {S_0 \lambda \over 2}\int {\bf v}\cdot \dot{\bf R}d^2x
\label{seven}
\end{equation}
where we have used the relation: ${\partial \phi \over \partial z} 
= {\partial \phi \over \partial {\bf R}}\dot{{\bf R}}, 
{\partial \phi \over \partial{\bf R}} = -\nabla\phi$ 
with $\dot {\bf R} \equiv {d{\bf R} \over dz}$. 
This can be obtained by the ``Euler-Lagrange'' equation for ${\bf R}$, 
which gives the ``balance of forces'' 
\begin{equation}
{\bf F}_C \equiv 
{d \over dz}{\partial L_C \over \partial \dot{\bf R}} - 
   {\partial L_C \over \partial {\bf R}} 
    = - {\partial \tilde H \over \partial {\bf R}}.
\label{eight}
\end{equation}
Using eq.(\ref{seven}), we get 
\begin{equation}
{S_0\lambda \over 2}\sigma \left({\bf k} 
\times \dot{\bf R}\right) = - {\partial \tilde H \over \partial {\bf R}}
\label{nine}
\end{equation}
where the ${\bf k }$ is the unit vector perpendicular 
to the $xy$-plane. Here $\sigma$ is defined as 
\begin{equation}
\sigma = \int_{R^2} \left({\partial v_y \over \partial x} 
    - {\partial v_x \over \partial y}\right)d^2x.
\label{ten}
\end{equation}
In deriving (\ref{nine}), we have used the relation 
${\partial v_x \over \partial X} = -{\partial v_x \over \partial x}$. 
The integrand of $\sigma$ is nothing but the vorticity which 
we put $\omega$. Using the expression for the 
velocity field in Eq.(\ref{seven}), we can write $\omega$ 
in terms of the angular functions: $\omega = (\nabla \times {\bf v})_z = 
 \sin\theta(\nabla\theta \times \nabla\phi)$, or in terms of the 
spin field ${\bf m}$
\begin{equation}
(\nabla \times {\bf v})_z = {\bf m}\cdot\left({\partial{\bf m} \over 
\partial x}\times{\partial {\bf m} \over \partial y}\right).
\label{eleven}
\end{equation}
The equation (\ref{eleven}) is an optical counterpart of 
a topological invariant of hydrodynamical origin \cite{Lamb}, 
which is written as 
\begin{equation} 
\sigma = \int_S \sin\theta d\theta\wedge d\phi
\label{thirt}
\end{equation}
where $S$ stands for the area in the pseudo-spin space 
$(\theta, \phi)$. $\sigma$ has a topological meaning, 
which depends on the boundary condition for $\theta(r)$. 
Namely, for the case a) corresponding to the boundary condition 
$\theta(\infty) = \pi$, the vortex configuration gives 
the mapping from the 
 compactified two-dimensional space $R^2 \cup {\infty} \simeq S^2$ 
to the pseudo-spin space $S_2$. Hence $\sigma$ in (\ref{thirt}) has 
a meaning of the degree of mapping for $S_2 \rightarrow S_2$ 
leading to the topological invariant $\sigma$: $\sigma= n$ ($n$=integer). 
For the case b) corresponding to the boundary condition 
$\theta(r) = {\pi \over 2}$, the mapping becomes $S_2 \rightarrow S_2/2 
({\rm hemisphere})$, so we have the topological invariant 
$\sigma= n/2$. The appearance of such two types of topological 
invariant is characteristics of a new type vortex presented here.

\bigskip
The authors would like to thank Mr. Akio Yoshimoto for his 
assisting the numerical solution. This work was carried out under 
the auspice of the Ritsumeikan University research grant.

%%
%Figures
%%
\newpage

\begin{thebibliography}{10}
\bibitem{Chiao}
R. Y. Chiao, E. Gamire, and C. H. Townes, 
Phys. Rev. Lett. {\bf 13}, 479 (1964). 
\bibitem{Snyder1}
A. W. Snyder, L. Polarian, and D.J. Mitchell, 
Opt. Lett. {\bf 17}, 789 (1992).
\bibitem{Swartz}
G. A. Swartzlander, Jr. and C. T. Law, 
Phys. Rev. Lett. {\bf 69}, 2503 (1992). 
\bibitem{Rozas}
D. Rozas, Z.S. Sacks and G.A. Swartzlander, Jr.,
Phys. Rev. Lett. {\bf 79}, 3399 (1997).
\bibitem{Snyder2}
A.~W. Snyder, S.~J. Hewlett and D.~J. Mitchell, Phys.~Rev. ~Lett.{\bf 72}
1012(1994). 
\bibitem{Born}
M. Born and E. Wolf, {\it Principle of Optics} 
(Pergamon, Oxford, 1975).
\bibitem{Landau}
L. Landau and E. Lifschitz, {\it Electrodynamics in Continuous Media}, 
chapter 11, Course of Theoretical Physics Vol.8 (Pergamon Oxford, 1968).
\bibitem{Kakigi}
H. Kuratsuji and S. Kakigi, Phys. Rev. Lett. 
{\bf 80}, 1888 (1998) and references cited therein. 
\bibitem{Landau2}
We here discard the term that is 
proportional to $\nabla\hat\epsilon$, since $\hat\epsilon$ 
is a slowly varying function. See \cite{Landau}.
\bibitem{Akhmanov} 
See S.A.Akhmanov, {\it Physical Optics}, Chapter 14
(Clarendon press, Oxford 1997), where the quasi-otical approximation 
is used ; $\left\vert{\partial^2{\bf f} \over \partial z^2 }\right\vert 
\ll k\left\vert{\partial{\bf f} \over \partial z}\right\vert$. 
\bibitem{Brosseau}
C.Brosseau, {\it Fundamental of Polarized Light: A Statistical 
Optics Approach}, Chapter 3, (John Wiley, New York, 1998).
\bibitem{Ono}
H. Ono and H. Kuratsuji, Phys. Lett. {\bf 186A}, 255(1994), 
H. Kuratsuji and H. Yabu, J. Phys. A{\bf 29}, 6505(1996).
\bibitem{Note}
We note that $\sigma$ in (\ref{ten}) does not depend on ${\bf R}$, 
because the integrand of $\sigma$ is a function of ${\bf x} - {\bf R}$, 
and with a change of the variable ${\bf x} \rightarrow {\bf x} - {\bf R}$, 
$\sigma$ becomes independent from ${\bf R}$.
\bibitem{Lamb}
e.g. H. Lamb, {\it Hydrodymanics}, (Cambridge University Press, 1932) p248. 
\end{thebibliography}
\end{document}